\documentclass[hyper]{JHEP} 

\usepackage{epsfig}

\newcommand\fverb{\setbox\pippobox=\hbox\bgroup\verb}

\newcommand\fverbdo{\egroup\medskip\noindent%

            \fbox{\unhbox\pippobox}\ }

\newcommand\fverbit{\egroup\item[\fbox{\unhbox\pippobox}]}

\newbox\pippobox

\title{Note About Hamiltonian Dynamics of String
in Non-Relativistic D3-Brane
Background}
\author{by J. Kluso\v{n}\\
     Department of Theoretical Physics and Astrophysics\\
                   Faculty of Science, Masaryk University\\
Kotl\'{a}\v{r}sk\'{a} 2, 611 37, Brno\\
Czech Republic\\
    E-mail: \email{klu@physics.muni.cz}}
\preprint{0909.5273}

 \abstract{This paper is devoted to the
study of the Hamiltonian dynamics of string
in non-relativistic D3-brane background. We discuss
different gauge fixing functions and construct
corresponding gauge fixed Hamiltonians.}


\def \bp{\mathbf{p}}

\def\pb #1{\left\{#1\right\}}
\def\bx{\mathbf{x}}

\def\mH{\mathcal{H}}

\def\mL{\mathcal{L}}

\newcommand \partt{\partial_\tau}
\newcommand \parts{\partial_\sigma}
\begin{document}
\section{Introduction and Summary}\label{first}
The AdS/CFT correspondence relates
conformal field theories in
$d$-dimensional flat space-times to
gravitational theories (super strings)
in asymptotically $AdS_{d+1}$
spacetimes
\cite{Maldacena:1997re,Gubser:1998bc,Witten:1998qj}.
Recently this correspondence has been
generalized to the description of some
 non-relativistic strongly
coupled conformal systems \footnote{For
review and extensive list of
references, see
\cite{Hartnoll:2009sz,McGreevy:2009xe}.}.
Non-relativistic conformal symmetry
contains the scaling transformation
\begin{equation}\label{scalrel}
x'^i=\lambda x^i \ , \quad t'=\lambda^z t \ ,
\end{equation}
where $z$ is a dynamical exponent. In
case of $z=2$ this symmetry is enhanced
to Schor\"{o}dinger symmetry
\cite{Hagen:1972pd,Niederer:1972zz,Henkel:1993sg,
Duval:1990hj,Duval:2009vt,Mehen:1999nd,Son:2005rv,
Leiva:2003kd,Correa:2008bi,Nishida:2007pj}.
It is remarkable that it is possible to
find the gravity dual of these
non-relativistic field theories
\cite{Son:2008ye,Balasubramanian:2008dm}
\footnote{For another solutions that
should be dual to non-relativistic
field theories, see
 \cite{Goldberger:2008vg,Barbon:2008bg,Herzog:2008wg,
Maldacena:2008wh,Adams:2008wt,Kovtun:2008qy,Hartnoll:2008rs,
Schvellinger:2008bf,Mazzucato:2008tr,Rangamani:2008gi,
Alishahiha:2009nm,Martelli:2009uc,Ross:2009ar,Kachru:2008yh,
Blau:2009gd,Yamada:2008if,
Bobev:2009mw,Bobev:2009zf,
Bagchi:2009my,Danielsson:2009gi,Bertoldi:2009vn,
Bertoldi:2009dt,Bertoldi:2009dt,Wen:2008hi,Nakayama:2008qm,Minic:2008xa,
Duval:2008jg,Akhavan:2008ep,Sakaguchi:2009de,Bagchi:2009ke,Alishahiha:2009np,
Bagchi:2009ca,
Volovich:2009yh,Akhavan:2009ns,Donos:2009en}.}.
The asymptotic metric in this case
reads
\begin{equation}\label{schro}
ds^2=\frac{R^2}{r^2}
(-\frac{dt^2}{r^{2(z-1)}}+ 2dt d\xi +
(dx^i)^2)+\frac{R^2}{r^2} dr^2
 +ds_M^2 \ ,
\end{equation}
where $R$ is characteristic radius of
space-time,  $\xi\sim \xi+L$ is a
compact light-like coordinate and where
$ ds^2_M$ is the metric of an
appropriate compact manifold which
allows (\ref{schro}) to be a solution
to the supergravity equations of
motion. Since $\xi$ is compact the
associated quantum number is
interpreted as the particle number.

As we said above $z$
that appears in the scaling
relation (\ref{scalrel})is a critical
exponent.
 The usual AdS case
corresponds to $z =1$. On the other
hand non-relativistic Dp-brane
backgrounds are characterized by
dynamical exponents $z=2$
\cite{Maldacena:2008wh,Mazzucato:2008tr}.
The backgrounds with more general values of $z$
were also found however it is still an
open question  of their consistency,
for nice discussion, see
\cite{Blau:2009gd}. For example, while
Schr\"{o}dinger space-time is
geodetically complete at $r\rightarrow
0$ for all $z\geq 1$, for $ z>  1$ the
detailed behavior  of geodesics near
$r = 0$ differs somewhat from the case
$z=1$ since ($r = 0$ is harder to
reach), and this may well have
implications for holography and, in
particular, for an appropriate notion
of boundary  in this context.

Since the holographic relation
between non-relativistic theory on
boundary and the string theory in
the bulk  is not
complete developed it is certainly
useful to
implement  ideas
 from standard $AdS/CFT$
correspondence  to this case in order
to gain insight into the subject
\footnote{Very nice discussion of the
non-relativistic holography can be
found in \cite{Nakayama:2009ed}}. For
example, it is well known that the
dynamics of the string in  uniform
gauge where momentum along compact
direction is uniformly distributed on
the string is  useful tool for the
study of the relations  between
observables in CFT on the boundary and
the string states in the bulk of
$AdS_5$
\cite{Arutyunov:2004yx}, for recent
review, see \cite{Arutyunov:2009ga}.
Since
the extremal non-relativistic background is
characterized by an
existence of compact light-like
direction $\xi$ and   off-diagonal
 metric component $g_{t\xi}$ together with an
absence of $g_{\xi\xi}$   we mean that it makes
sense  to study the Hamiltonian dynamics of the
string in this background where different
gauge fixing conditions are imposed.
 We will see  that in the standard gauge when
 the world-sheet time coincides with target
 space time the Hamiltonian dynamics is
well  defined on condition of non-zero
momentum along $\xi$ direction. More
precisely, we consider two gauge
fixing: The first one when we demand
that the world-sheet coordinates
coincide with target space ones and the
second one when we demand that the
world-sheet time coincides with the
background time and when the momentum
$P_\xi$ is uniformly distributed along
the world-sheet of the string. In both
these cases we find that  gauge fixed
Hamiltonians have manifestly
non-relativistic form. We also argue
that in the gauge when
the string wraps
$\xi$ direction there is no pure time
dependent solutions of the equations of
motion. On the other hand
in case of uniform gauge there is a possibility
of pure time dependent dynamics and
we find that the string with finite
energy  cannot reach the boundary of
space time at $r=0$  which is in agreement
with previous analysis of radial geodesics in
Schr\"{o}dinger spacetimes  \cite{Blau:2009gd}.
Finally we construct the string in uniform gauge
in the non-extremal non-relativistic
background and we find the
Hamiltonian density has "square-root"
form as in case of relativistic background.

This paper can be extended in many
directions. For example, it will be interesting
to study the static solutions corresponding
to Wilson lines. We can also search for more
general configurations of the string
 as for example
folded or pulsating strings  with
analogy with solutions found in the
context of AdS/CFT correspondence (For
review, see
\cite{Tseytlin:2003ii,Tseytlin:2004xa}).

This paper is organized as follows. In
 section (\ref{second}) we review
the Hamiltonian dynamics of the bosonic
string in general background. In
section (\ref{third}) we consider the
static gauge for the bosonic string in
non-relativistic background and we find
the gauge fixed form of the Hamiltonian
density. In section (\ref{fourth}) we
study the uniform gauge
 and we also study the dynamics
of the radial mode. Finally in section
(\ref{fifth}) we find the uniform gauge
fixed form of the Hamiltonian for the
string in non-extremal non-relativistic
D3-brane background.

\section{Hamiltonian Formalism for
String in Non-Relativistic D3-brane
Background}\label{second} In this
section we review
the construction of the
Hamiltonian for the bosonic string in
general background. Recall that the
action for the bosonic string in general
background takes the form
\begin{equation}\label{Polact}
S=-\frac{1}{4\pi\alpha'} \int d^2\sigma
\sqrt{-\gamma}\left(\gamma^{\alpha\beta}
g_{MN}\partial_\alpha x^M\partial_\beta x^N
-\epsilon^{\alpha\beta}b_{MN}\partial_\alpha
x^M\partial_\beta x^N\right) \ ,
\end{equation}
where $\gamma_{\alpha\beta}$ is a
two-dimensional world-sheet metric, $
\sigma^\alpha \ , \alpha,\beta=0,1 \ ,
 \sigma^0=\tau \ ,
\sigma^1=\sigma$ are
world-sheet coordinates and
 $\epsilon^{\alpha\beta}=
\frac{\varepsilon^{\alpha\beta}}{\sqrt{-\gamma}}
\ ,
\varepsilon^{\tau\sigma}=-\varepsilon^{\sigma
\tau}=1$. Further, $x^M\ ,
M,N=0,\dots,9$ are modes that
parameterize the embedding of the
string into target space-time with
background metric $g_{MN}$ and NS
two-form $b_{MN}$.

Our goal is to develop
 the Hamiltonian
formalism from the action
(\ref{Polact}). It is
convenient to use following
parametrization of  the world-sheet
metric
\begin{equation}
\gamma_{\alpha\beta}= \left(
\begin{array}{cc}
-N^2_T+\omega N_S^2 & \omega N_S \\
\omega N_S & \omega \\
\end{array}\right) \
\end{equation}
so that
\begin{equation}
\det \gamma=-N_T^2 \omega \ , \quad
\gamma^{\alpha\beta}=
\left(\begin{array}{cc}
-\frac{1}{N^2_T}
 & \frac{N_S}{N^2_T} \\
 \frac{N_S}{N^2_T}
 &
 \frac{1}{\omega}-\frac{N_S^2}{N_T^2}
 \\ \end{array}\right) \ .
\end{equation}
With this form of the world-sheet
metric the action
(\ref{Polact}) takes the form
\begin{eqnarray}\label{Polact2}
S&=&\frac{1}{4\pi\alpha'} \int
d^2\sigma N_T\sqrt{\omega} (g_{MN}\nabla_\tau
x^M\nabla_\tau
x^N-\frac{1}{\omega}
g_{MN}\parts x^M\parts x^N+\nonumber \\
&+& b_{MN}
\partt x^M\parts x^N-b_{MN}
\parts x^M\partt x^N) \ , \nonumber \\
 \nonumber \\
\end{eqnarray}
where
\begin{equation}
\nabla_\tau x^M=\frac{1}{N_T} (\partt
x^M- N_S\parts x^M) \ .
\end{equation}
Now we introduce the momenta
$\pi_T,\pi_S,\pi_\omega$
conjugate to $N_T,N_S$ and $\omega$
with canonical Poisson brackets
\begin{equation}
\pb{N_T(\sigma),\pi_T(\sigma')}=
\delta(\sigma-\sigma') \ , \quad
\pb{N_S(\sigma),\pi_S(\sigma')}=
\delta(\sigma-\sigma') \ , \quad
\pb{\omega(\sigma),
\pi_\omega(\sigma')}=\delta(\sigma-
\sigma') \ .
\end{equation}
Then due to the fact that the action
(\ref{Polact2}) does not contain the
time derivative of metric we find that
these momenta are primary constraints
of the theory
\begin{equation}\label{pricon}
\pi_T=\frac{\delta S}{\delta \partt
N_T}\approx 0 \ , \quad
\pi_S=\frac{\delta S}{\delta \partt
N_S}\approx 0 \ , \quad
\pi_\omega=\frac{\delta S}{\delta
\partt \omega}\approx 0 \ .
\end{equation}
As the next step we introduce the
momenta $p_M$  conjugate to $x^M$ so
that $\pb{x^M(\sigma),p_N(\sigma')}=
\delta^M_N\delta(\sigma-\sigma')$. From
(\ref{Polact2}) we get
\begin{equation}
p_M=\frac{\delta S}{\delta
\partt x^M}=\frac{1}{2\pi\alpha'}( \sqrt{\omega}
g_{MN}\nabla_\tau x^N+b_{MN}\parts x^N)
\ .
\end{equation}
Then it is easy to find  the
Hamiltonian density  in the form
\begin{eqnarray}\label{H}
\mH=\partt x^Np_M-\mL=
\frac{N_T}{\sqrt{\omega}} \mH_T+N_S \mH_S
\ , \nonumber \\
\end{eqnarray}
where
\begin{eqnarray}
 \mH_T&=&\frac{1}{4\pi\alpha'}
\left((2\pi\alpha'p_M-b_{NK}\parts x^K)
g^{MN} (2\pi\alpha'p_N-b_{NL}\parts
x^L)+
\parts x^M
g_{MN}\parts x^N\right) \ , \nonumber \\
\mH_S&=&\parts x^Np_N \ . \nonumber \\
\end{eqnarray}
Using the  Hamiltonian (\ref{H})
we easily determine
the time evolution of the primary
constraints as
\begin{eqnarray}
\partt \pi_T&=&\pb{\pi_T,H}=
-\frac{1}{\sqrt{\omega}}\mH_T \ ,
\nonumber \\
\partt \pi_\omega&=&\pb{\pi_\omega,H}=
\frac{N_T}{\omega^{3/2}}\mH_T
 \ ,
\nonumber \\
\partt \pi_S&=&\pb{\pi_S,H}=
-\mH_S  \ .  \nonumber \\
\end{eqnarray}
Since the  constraints (\ref{pricon})
 have to be
preserved during the time evolution of
the system the equations above imply an
existence of the  secondary constraints
\begin{equation}\label{secondco}
\mH_T\approx 0 \ , \quad \mH_S\approx 0
\ .
\end{equation}
Finally it can be  shown that the
constraints $\mH_T$ and $\mH_S$ form
the closed set of constraints and the
consistency of their time evolution
does not impose additional ones. In
summary, we have collections of the
first class constraints $(\pi_T\approx
0 \ , \pi_S\approx 0 ,
\pi_\omega\approx 0 \ ,  \mH_T\approx 0
\ , \mH_S\approx 0)$ and consequently
the total Hamiltonian density is equal
to \footnote{For reviews of the
dynamics of constrained systems, see
\cite{Govaerts:2002fq,Govaerts:1991gd,Henneaux:1992ig}.}
 \begin{equation}\label{ungaugedH}
\mH=\left(\frac{N_T}{\sqrt{\omega}}+f_T\right)
\mH_T+(N_S+f_S)\mH_S+ u_T \pi_T+
u_S\pi_S+u_\omega \pi_\omega\approx 0 \ ,
\end{equation}
where $f_T,f_S,u_T,u_S,u_\omega$ are
arbitrary
functions that generally depend on
world-sheet coordinates and on the
phase-space variables. We will see
in next sections that
these functions are uniquely specified
when we introduce the gauge fixing
functions that impose additional
constraints on the theory.

In what follows we concentrate on the
dynamics of the bosonic string in
non-relativistic D3-brane background
\cite{Mazzucato:2008tr}
\footnote{Clearly this analysis can be
generalized to more general
form of the non-relativistic background.}
\begin{eqnarray}\label{met}
ds^2&=&\frac{R^2}{r^2}
\left(-\frac{dt^2}{r^2}+ 2dt d\xi +
(dx^i)^2\right)+\frac{R^2}{r^2} \left(dr^2+r^2
((d\chi+\mathcal{A})^2+ds_{P^2}^2)\right) \ ,
\nonumber
\\
B&=&\frac{R^2}{r^2}(d\chi+\mathcal{A})
\wedge dt \ , \nonumber \\
\end{eqnarray}
where
\begin{eqnarray}\label{dsP}
\mathcal{A}&=&\frac{1}{2}\sin^2\mu
\sigma_3 \ ,
\nonumber \\
ds^2_{P^2}&=&d\mu^2+\frac{1}{4}\sin^2\mu
(\sigma_1^2+\sigma^2_2+
\cos^2\mu \sigma^2_3) \  ,  \nonumber \\
\end{eqnarray}
and where
 $\sigma_i$ are $SU(2)$
left-invariant forms that satisfy
\begin{equation}
d\sigma_i=-\frac{1}{2}\epsilon_{ijk}
\sigma_j\wedge \sigma_k \ .
\end{equation}
Explicitly they have the form
\begin{eqnarray}
\sigma_1&=&\cos\psi d\theta+\sin\theta
\sin \psi d\phi \ , \nonumber \\
\sigma_2&=&-\sin\psi d\theta+\sin\theta
\cos\psi d\phi \ , \nonumber \\
\sigma_3&=& d\psi +\cos\theta d\theta \
.
\nonumber \\
\end{eqnarray}
For simplicity we introduce the
symbol $y^a,a=1,2,3,4$
defined as
$y\equiv
(\mu,\psi,\theta,\phi)$ and
write
the line element of squashed $S^5$ as
\begin{equation}\label{defsphere}
ds^2_{S^5}= d\chi^2+g_{a\chi} dy^a
d\chi+g_{\chi a}d\chi dy^a+ g_{ab} dy^a
dy^b \ ,
\end{equation}
where the explicit form of metric
components can be easily determined
from (\ref{met}) and (\ref{dsP}).
 In the same way we write
\begin{eqnarray}\label{defb}
b_{\chi
t}=-b_{t\chi}=\frac{1}{2}\frac{R^2}{r^2}
\ , \quad  b_{at}=-b_{ta}= \frac{1}{2}
\frac{R^2}{r^2} \mathcal{A}_a \ .
\nonumber
\\
\end{eqnarray}
Now we are ready to proceed to the
construction of the gauge-fixed form of
the bosonic string  action on
non-relativistic background. Recall
that the Hamiltonian (\ref{ungaugedH})
is a collection of the first class constraints
and hence vanishes on constraint surface.
The gauge fixing of the first class
constraints consists in introduction
of additional constraints in the theory
that have non-trivial Poisson brackets
with the initial first class constraints
on the constraint surface.
 Then the
collections of the original first class
constraints together with the gauge
fixing functions form the set of the
second class constraints. The
characteristic property of the second
class
constraints is that they can be explicitly
solved and hence  some degrees of freedom
can be eliminated.
 It is
also clear that we should replace the
original Poisson brackets with the
Dirac ones. However it turns out that
for the gauge fixed functions that we
use in
this paper the Poisson brackets
coincide with the Dirac ones.

Of course all these comments are valid for
any theory with the  first class constraints.
 However the crucial property of
 the extremal
non-relativistic background  is an
absence of the metric component
$g_{\xi\xi}$. Then  using the form
of the metric (\ref{met}) we find
\begin{equation}
g^{tt}=0 \ , \quad g^{\xi\xi}=
-\frac{g_{tt}}{g^2_{t\xi}} \ , \quad
g^{t\xi}=\frac{1}{g_{t\xi}} \ .
\end{equation}
The fact that $g^{tt}$ is zero will
have an important consequence
for
the construction of the gauge fixed
Hamiltonian.
\section{Static Gauge}\label{third}
In this section we study the gauges
when world-sheet coordinates coincide with
some target space coordinates. We consider
two situations:
\subsection{Static Gauge
$t=\tau,r=\sigma$} This gauge is
defined  by introducing
 following two gauge
fixing functions
\begin{equation}\label{gff}
G_1:t-\tau=0 \ , \quad G_2:r-\sigma=0 \
.
\end{equation}
Let us now show that these functions
fix all gauge freedom of the theory.
We firstly  determine
the Poisson brackets between $G_1,G_2$
and $\mH_T,\mH_S$. For the background
(\ref{met}) these constraints take the
form
\begin{eqnarray} \mH_S&=&p_t\parts
t+p_r\parts r+ \mH^{r.s.g.}_S \ ,
\nonumber \\
\mH_T
 &=&\frac{1}{4\pi\alpha'}
 (2(2\pi\alpha')^2 p_t g^{t\xi}p_\xi
 +  (2\pi\alpha')^2 p_r g^{rr}p_r+
  g_{rr}\parts
r\parts r+
  \nonumber \\
  &+&g_{tt}\parts t\parts t+2g_{t\xi}\parts
  t\parts \xi)+\mH^{r.s.g.}_T \ ,
  \nonumber \\
\end{eqnarray}
where
\begin{eqnarray}
\mH^{r.s.g.}_T&=&\frac{1}{4\pi\alpha'}
((2\pi\alpha')^2p_\xi g^{\xi\xi}p_\xi
-2(2\pi\alpha')^2(b_{ta}\parts y^a+
 b_{t\chi}\parts \chi ) g^{t\xi}
  p_\xi
+\nonumber \\
&+&
(2\pi\alpha')^2p_{\chi}g^{\chi\chi}p_{\chi}+
2 (2\pi\alpha')^2 p_{\chi}g^{\chi
a}p_a+ (2\pi\alpha')^2 p_a g^{ab}p_b+
\nonumber \\
&+&\frac{1}{4\pi\alpha'} (g_{ij}
\parts x^i\parts x^j+
g_{\chi\chi}(\parts \chi)^2+2g_{a\chi}
\parts \chi \parts y^a+
g_{ab}\parts y^a\parts y^b) \ ,
\nonumber \\
\mH^{r.s.g.}_S&=& p_\xi\parts \xi+
p_{\chi}\parts \chi+ p_a\parts y^a+
p_i\parts x^i \ , \nonumber \\
\end{eqnarray}
and where for simplicity of notation
we ignore from the beginning
terms proportional to $b_{\xi t}\parts t \ , b_{a t}\parts t$ that vanish in
all gauges studied in this paper.
It is easy to determine the Poisson
brackets between $G_1,G_2$ and
$\mH_T,\mH_S$
\begin{eqnarray}\label{G11H}
\pb{G_1(\sigma),\mH_T(\sigma')}&=&2\pi\alpha'
g^{t\xi}p_\xi(\sigma)\delta(\sigma-\sigma')
\ , \nonumber \\
\pb{G_1(\sigma),\mH_S(\sigma')}&=&
\parts
t(\sigma)\delta(\sigma-\sigma')\approx
0 \ , \nonumber \\
\pb{G_2(\sigma),\mH_T(\sigma')}&=&
2\pi\alpha'g^{rr}p_r(\sigma)\delta(\sigma-\sigma') \ ,
\nonumber \\
\pb{G_2(\sigma),\mH_S(\sigma')}&=&
\parts r(\sigma)\delta(\sigma-\sigma')\approx
\delta(\sigma-\sigma')
 \ ,
\nonumber \\
\end{eqnarray}
where $\approx 0$  means that these
results hold on the constraint surface
$G_1=G_2=\mH_T=\mH_S=0$. Clearly the
matrix of Poisson brackets (\ref{G11H})
is
non-singular and hence the collection
of the original
 first class constraints with the gauge
 fixed functions
forms the system of
the second class constraints. It
is also easy to see that the Dirac
brackets of the canonical variables on
the reduced space  coincide with the
Poisson brackets.

Clearly we also have to fix the gauge
freedom that is related to the
generators $\pi_S,\pi_T$ and
$\pi_\omega$ and we do it by imposing
the condition
\begin{equation}\label{gffm}
G_{\pi_T}:N_T-1=0 \ , \quad  G_{\pi_S}: N_S=0
\ , \quad G_{\pi_\omega}:\omega-1=0
\ .
\end{equation}
It is easy to see  that Poisson
brackets of these gauge fixed functions
with the constraints $\pi_T,\pi_S,
\pi_\omega$ are non-zero and again the
collections of gauge fixed functions
given in (\ref{gffm}) together with the
primary constraints (\ref{pricon}) form
the second class constraints.

We can also see that the functions
(\ref{gff}) fix the gauge from the
following arguments. The consistency of
the theory implies that the conditions
(\ref{gff})  have to be preserved
during the time evolution of the
system. Since the time evolution of any
phase space function is governed by the
equation
\begin{equation}
\frac{d\Phi}{d\tau}=
\partt \Phi+\pb{\Phi,H} \
\end{equation}
we find
\begin{eqnarray}
\frac{dG_1}{d\tau}&=&-1+(\frac{N_T}{\sqrt{\omega}}+f_T)
2\pi\alpha'
g^{t\xi}p_\xi+(N_S+f_S)\parts t \approx
\nonumber \\
&\approx & -1+(1+f_T) 2\pi\alpha'
g^{t\xi}p_\xi=0 \nonumber \\
\end{eqnarray}
and hence $f_T$ is equal to
\begin{equation}\label{fT}
f_T=\frac{1-2\pi\alpha' g^{t\xi} p_\xi
}{2\pi\alpha' g^{t\xi}p_\xi} \ .
\end{equation}
 This result also implies that the
gauge fixed theory is well defined on
condition that $p_\xi\neq 0$.  On the
other hand the requirement of the
preservation of $G_2$ under the time
evolution implies
\begin{eqnarray}
\frac{dG_2}{d\tau}&=&
(\frac{N_T}{\sqrt{\omega}}+f_T)
2\pi\alpha'g^{rr}p_r+(N_S+f_S)\parts r
\approx
\nonumber \\
&\approx &(1+f_T)2\pi\alpha' g^{rr}p_r
+f_S=0 \nonumber \\
\end{eqnarray}
and with the help of (\ref{fT}) we get
$f_S=-\frac{g^{rr}}{g^{t\xi}}\frac{p_r}{p_\xi}=
-\frac{p_r}{p_\xi}$. Finally the
preservation of the gauge fixing
functions (\ref{gffm})
 implies that
\begin{equation}
u_{\pi_T}=0 \ , \quad
u_{\pi_S}=0 \ , \quad  u_{\pi_\omega}=0 \ .
\end{equation}
To conclude, the result of the gauge fixing
procedure is that the Hamiltonian
density is strongly zero and that the original
constraints together with the gauge
fixing functions consist a collection
of the second class constraints. Then
solving the constraint $\mH_S=0$ we
find
\begin{equation}\label{prsg}
p_r=-\mH^{r.s.g.}_S \ .
\end{equation}
Further, the Hamiltonian density of
the  gauge fixed theory is equal
to $\mH^{r. s. g.}=-p_t$. This identification
can be nicely seen from the fact that the
action of gauge fixed theory is
\begin{equation}
S=\int d^2\sigma (\partt x^Mp_M-\mH)=
\int d^2\sigma (p_a\partt y^a+p_i\partt
x^i+ p_\xi \partt \xi+p_\chi\partt
\chi-p_t) \ ,
\end{equation}
where we used the fact that the
original Hamiltonian density $\mH$
strongly vanishes. The Hamiltonian
density $\mH^{r.s.g.}$ can be derived
by solving the constraint $\mH_T=0$
\begin{eqnarray}
\mH^{r.s.g.}&=&-p_t= \frac{1}{8\pi^2
\alpha'^2 g^{t\xi}p_\xi} \left(
(2\pi\alpha')^2 p_r g^{rr}p_r
+g_{rr}+\mH^{r.s.g.}_T\right)=
\nonumber \\
&=& \frac{1}{8\pi^2 \alpha'^2
g^{t\xi}p_\xi} \left( (2\pi\alpha')^2
(\mH_S^{r.s.g.})^2 g^{rr}
+g_{rr}+\mH^{r.s.g.}_T\right) \ ,
\nonumber \\
\end{eqnarray}
where in the second step we used
(\ref{prsg}).

Clearly the static gauge can be also
imposed  in the Lagrangian formalism
of the bosonic string however we mean
that it was  instructive to perform this
analysis  in the Hamiltonian
formalism in order to introduce main
ideas and notations.
Note also that this Hamiltonian is well
defined on condition that $p_\xi\neq
0$.  We also see that the static gauge
can be implemented for either an open
string or for an infinite long string
since the variable $r$ is non-periodic.
In fact, string in  the gauge
$t=\tau, r=\sigma$ can be used
for the description of the
the Wilson
loops in non-relativistic theories
as in case of AdS/CFT correspondence
\cite{Maldacena:1998im,Rey:1998bq,Rey:1998ik}.
\subsection{Static Gauge $t=\tau \ ,
\xi=\sigma$}
As in  previous section we
 fix the primary constraints
(\ref{pricon}) with functions
 (\ref{gffm}). Then we
fix the constraints $\mH_T\approx 0 \ ,
\mH_S\approx 0$ by introducing
 two conditions
\begin{equation}\label{sgxi}
G_t: t-\tau=0 \ , \quad G_\xi:
\xi-m\sigma=0 \ ,
\end{equation}
where the number
 $m$ has following interpretation.
We known that $\xi$ is a compact variable
with the identification $\xi\sim \xi
+L$. As a consequence
the gauge (\ref{sgxi})  is
defined for closed string where
$\sigma$ belongs to the finite interval
$\sigma\in[-r,r]$ and all world-sheet
modes are periodic on given interval.
 The number $r$ can
be determined from the gauge fixing
condition (\ref{sgxi}) since
\begin{equation}
\xi(r)-\xi(-r)= \int_{-r}^r d\sigma
\parts \xi= \int_{-r}^r d\sigma m =2r
m=mL
\end{equation}
and we see that for $r=\frac{1}{2}L$
the number $m$ is the wrapping number
that counts how many times the string
wraps the compact direction $\xi$.

In order to simplify the final results
we will write the constraints
$\mH_S,\mH_T$ in the form
\begin{eqnarray}\label{mHSTxi}
\mH_S&=&p_t\parts t+p_\xi\parts \xi+
\mH^{\xi.g.f.}_S \ , \nonumber \\
\mH_T&=& \frac{1}{4\pi\alpha'}
 (4\pi\alpha' (2\pi \alpha' p_t-b_{tI}\parts \bx^I ) g^{t\xi}
  p_\xi+
  (2\pi\alpha')^2 p_\xi g^{\xi\xi}p_\xi+
  \nonumber \\
  &+&g_{tt}\parts t\parts t+2g_{t\xi}\parts
  t\parts \xi)+\mH^{\xi.g.f.}_T \ ,
  \nonumber \\
\end{eqnarray}
where
\begin{eqnarray}
\mH^{\xi.g.f.}_S&=& p_I\parts \bx^I \ ,
\quad
\mH^{\xi.g.f.}_T=\frac{1}{4\pi\alpha'}
((2\pi\alpha')^2 p_I g^{IJ}p_J+
\parts \bx^I g_{IJ}\parts \bx^J) \ ,
\nonumber \\
\end{eqnarray}
where $\bx^I=(r,x^i,\chi,y^a)\ ,
I=1,\dots,8$ and components of the
metric $g_{IJ}$ and two form fields
$b_{tI}$ follow from (\ref{met}),
(\ref{defsphere}) and (\ref{defb}).
Note that we again ignore terms
$b_{It}\parts t$ that  vanish
in the gauge (\ref{sgxi}).

As usual we start with the
 calculation of
Poisson brackets of the gauge fixing
functions  (\ref{sgxi}) with
$\mH_T,\mH_S$
\begin{eqnarray}
\pb{G_t(\sigma),\mH_T(\sigma')}&=&
2\pi\alpha' g^{t\xi}p_\xi(\sigma)
\delta(\sigma-\sigma') \  , \nonumber
\\
\pb{G_t(\sigma),\mH_S(\sigma')}&=&\parts
t(\sigma)\delta(\sigma-\sigma')\approx
0 \nonumber \\
\pb{G_\xi(\sigma),\mH_T(\sigma')}&=&
 ((2\pi \alpha' p_t-b_{tI}\parts \bx^I ) g^{t\xi}
  +
 2 \pi\alpha' p_\xi g^{\xi\xi})(\sigma)
 \delta(\sigma-\sigma')\ , \nonumber \\
 \pb{G_\xi(\sigma),\mH_S(\sigma')}&=&
 \parts
 \xi(\sigma)\delta(\sigma-\sigma')\approx
 m \delta(\sigma-\sigma') \ .  \nonumber \\
\end{eqnarray}
Then the requirement that (\ref{sgxi})
are preserved during the time
evolutions of the system uniquely
determine the functions $f_T,f_S$ and
we find
\begin{eqnarray}
f_T&=&\frac{1-2\pi\alpha'
g^{t\xi}p_\xi}
{2\pi\alpha' g^{t\xi}p_\xi} \ ,  \nonumber \\
f_S&=&-\frac{1}{ 2\pi\alpha' m
g^{t\xi} p_\xi}
 ((2\pi \alpha' p_t-b_{tI}\parts \bx^I ) g^{t\xi}
  +
 2 \pi\alpha' p_\xi g^{\xi\xi})
 \ . \nonumber \\
\end{eqnarray}
Next steps are as in the previous
section.
 Solving $\mH_S=0$ we
express $p_\xi$ as functions
of the reduced phase space variables
$\bx^I$
\begin{equation}\label{pchimH}
p_\xi=-\frac{1}{m}\mH^{\xi.g.f.}_S \ .
\end{equation}
Solving the constraint $\mH_T=0$ we
find that the gauge fixed Hamiltonian
$\mH^{\xi.g.f.}=-p_t$ takes the form
\begin{eqnarray}\label{gfHxi}
\mH^{\xi.g.f.}=-\frac{1}{2\pi\alpha'}
b_{tI}\parts
\bx^I+\frac{1}{2}\frac{g^{\xi\xi}}{g^{t\xi}}p_\xi
+\frac{1}{2\pi\alpha'
g^{t\xi}p_\xi}\mH_T^{\xi.g.f.}=
\nonumber \\
  =-\frac{1}{2\pi\alpha'}
b_{tI}\parts \bx^I+
 \frac{1}{2m}\frac{g_{tt}}{g_{t\xi}}
 \mH^{\xi.g.f.}_S-\frac{m
 g_{t\xi}}{2\pi\alpha'}
 \frac{\mH^{\xi.g.f.}_T}{\mH^{\xi.g.f.}
 _S} \ , \nonumber \\
\end{eqnarray}
where in the final step we used
(\ref{pchimH}). Clearly  the
Hamiltonian (\ref{gfHxi}) is well
defined on condition that
 $\mH^{\xi.g.f.}_S\neq
0$. In particular, this result implies
that there do not exist pure time
dependent solutions.

It is also instructive to determine the
Lagrangian density from the gauge fixed
form of the Hamiltonian density
(\ref{gfHxi}). The first step is to
find time derivative of $\bx^I$ using
the canonical equation of motion
\begin{eqnarray}
\partt \bx^I=
\pb{\bx^I,H^{\xi.g.f.}}=
\frac{1}{2m}\frac{g_{tt}}{g_{t\xi}}
\parts \bx^I-m
g_{t\xi}
\frac{g^{IJ}p_J}{\mH^{\xi.g.f.}_S}+
\frac{m g_{t\xi}
}{2\pi\alpha'}\frac{\mH^{\xi.g.f.}_T}
{(\mH^{\xi.g.f.}_S)^2}\parts \bx^I \ .
\nonumber \\
\end{eqnarray}
Then after some algebra we find the
Lagrangian density in the form
\begin{eqnarray}\label{NGact}
\mL_{\xi.g.f.} &=&\parts \bx^I p_I-\mH^{\xi.g.f.}=
\frac{1}{2\pi\alpha'}
b_{tI}\parts \bx^I-\nonumber \\
 &-& \frac{1}{2\pi\alpha'}
\sqrt{- (g_{tt}+ (\partt \bx^I\partt
\bx_I))(\parts \bx^I\parts \bx_I))+(m
g_{t\xi}+\partt \bx^I \parts
\bx_I)^2} \ . \nonumber \\
\end{eqnarray}
In other words we reproduced the
Nambu-Goto action for bosonic string in
the gauge  $t=\tau \ , \quad
\xi=m\sigma$. The form of the Lagrangian
density (\ref{NGact}) again implies
that it is not possible
to find pure time dependent solution
of the equation of motion. It will be
interesting   to
find configuration that depends on
$\tau$ and $\sigma$ and that
is a  solution of the equation
of motion that follow from (\ref{NGact}).
We hope to return to this problem in
future.
\section{Uniform Gauge $t=\tau, p_\xi=\mathrm{const}$}
\label{fourth}
 In this section we
construct the Hamiltonian for the
bosonic string in the background
(\ref{met}) where the momentum $P_\xi$
is uniformly distributed  along the
string.  In fact, since the background
(\ref{met}) is invariant under the
shift $\xi'=\xi+\delta \xi$ with
$\delta \xi=\mathrm{const}$ the standard
arguments imply an existence
of   the conserved
charge
\begin{equation}
P_\xi=\int_{-r}^{r} d\sigma p_\xi
 \ ,
\end{equation}
where, as in previous section, we
consider the closed string with
$\sigma\in [-r,r]$.
Following the seminal work
\cite{Arutyunov:2004yx} we
 impose
the uniform gauge by introducing two
gauge fixing functions
\begin{equation}\label{gfu}
G_1:t-\tau=0 \ , \quad
G_2: p_\xi-\bp=0 \ ,
\end{equation}
where $\bp=\mathrm{const}$.
Since
\begin{eqnarray}\label{pbGH}
\pb{G_1(\sigma),\mH_T(\sigma')}&=&2\pi
\alpha' g^{t\xi}p_\xi
(\sigma)\delta(\sigma-\sigma')\approx
2\pi\alpha' g^{t\xi}\bp
\delta(\sigma-\sigma') \ ,
\nonumber \\
\pb{G_1(\sigma),\mH_S(\sigma')}&=&
\parts
t(\sigma)\delta(\sigma-\sigma')\approx
0 \ ,
\nonumber \\
\pb{G_2(\sigma),\mH_T(\sigma')}&=&
-\frac{1}{\pi\alpha'}
g_{t\xi}(\sigma')\partial_{\sigma'}t(\sigma')
\partial_{\sigma'}\delta(\sigma-\sigma')
\approx 0 \ ,
\nonumber \\
\pb{G_2(\sigma),\mH_S(\sigma')}&=&
-p_{\xi}(\sigma')
\partial_{\sigma'}\delta(\sigma-\sigma')\approx
-\bp
\partial_{\sigma'}\delta(\sigma-\sigma')
\
\nonumber \\
\end{eqnarray}
we find that (\ref{gfu}) together with
(\ref{secondco}) form the second class
constraints. Further the requirement of
the preservation of the gauge fixing
functions (\ref{gfu}) during the time
evolution of the system implies
following set of equations
\begin{eqnarray}\label{G1}
\frac{dG_1}{d\tau}&=& -1+
(\frac{N_T}{\sqrt{\omega}}+f_T)
2\pi\alpha' g^{t\xi} \bp \approx 0 \ ,  \nonumber \\
\frac{dG_2}{d\tau}&=&
-\left(\frac{1}{\pi\alpha'}
g_{t\xi}(\frac{N_T}{\sqrt{\omega}}+f_T)(r)
+\bp(N_S+f_S)(r)\right)
\delta(\sigma-r)+\nonumber
\\
&+&\left(\frac{1}{\pi\alpha'}
g_{t\xi}(\frac{N_T}{\sqrt{\omega}}+f_T)(-r)
+\bp(N_S+f_S)(-r)\right)
\delta(\sigma+r)+\nonumber \\
&+&\parts\left[\frac{1}{\pi\alpha'}
g_{t\xi}(\frac{N_T}{\sqrt{\omega}}+f_T)
+(N_S+f_S)\bp\right]\approx 0 \ ,
\nonumber \\
\end{eqnarray}
using  the Poisson brackets given in
(\ref{pbGH}). Fixing the metric
components as in (\ref{gffm})
 we find that the equations
 (\ref{G1}) imply
 \begin{eqnarray}
f_T=\frac{1}{2\pi\alpha' g^{t\xi}\bp}-1
\ , \quad f_S=-\frac{g_{t\xi}^2}{\bp^2}
\  .
\nonumber \\
\end{eqnarray}
Now with the help of the second gauge fixed function
(\ref{gfu})  we easily calculate the total conserved
momentum $P_\xi$
\begin{equation}
P_\xi=\int_{-r}^r d\sigma p_\xi= 2r \bp
\end{equation}
and hence we find the relation between
$r,\bp$ and $P_\xi$
\begin{equation}\label{rbpP}
 r=\frac{1}{2\bp}P_\xi \
\end{equation}
that will be useful below.

Let us proceed to the construction of
the gauge fixed Hamiltonian. Solving
the constraint $\mH_S=0$ we find
\begin{equation}\label{partxi1}
\parts \xi=-\frac{1}{\bp}
\mH_S^{\xi.g.f.} \ .
\end{equation}
 Since $\xi$ is a periodic variable
 string can wrap this
direction. In other words we should
identify the end points of the string
as
\begin{equation}
\triangle \sigma=\xi(r)-\xi(-r)=mL \ ,
\end{equation}
where $m$ is the winding number. Then
using (\ref{partxi1}) we can rewrite
this condition in the equivalent form
\begin{eqnarray}
& &\xi(r)-\xi(-r)=
\int_{-r}^r d\sigma
\parts \xi=\nonumber \\
&=&-\frac{1}{\bp} \int_{-r}^r d\sigma
\mH^{\xi.g.f.}_S=mL \ .
\nonumber \\
\end{eqnarray}
Clearly all classical solutions have to
obey this condition.
As the last step we determine the form
of the gauge fixed Hamiltonian density
$\mH^{p.g.f.}=-p_t$. Using the
constraint $\mH_T=0$ together with
(\ref{partxi}) we find that it takes
the form
\begin{eqnarray}\label{Hgfp}
\mH^{p.g.f.}
&=&-\frac{1}{2\pi\alpha'} b_{tI}\parts \bx^I
-\frac{\bp}{2}
\frac{g_{tt}}{g_{t\xi}}+\frac{g_{t\xi}}
{2\pi\alpha' \bp}\mH^{\xi.g.f.} \ . \nonumber
\\
\end{eqnarray}
We see that the Hamiltonian is positive
definite for $\bp>0$. Further, as follows from
 (\ref{rbpP})  we can set $\bp$
to be equal to  $\bp=\frac{1}{2\pi\alpha'}$
by rescaling $r$.

Let us now find the Lagrangian density from the Hamiltonian
density  (\ref{Hgfp}).
Since
\begin{eqnarray}
\partt \bx^I=\pb{\bx^I,H^{p.g.f.}}=
2\pi\alpha' g_{t\xi} g^{IJ}p_J \
\nonumber \\
\end{eqnarray}
and since $g^{IJ}$ is invertible
we can express $p_I$ as functions
of $\partt \bx,\bx$ and consenquently
the Lagrangian
density takes the form
\begin{eqnarray}\label{Lxif}
& &\mL^{p.g.f.}=
\partt \bx^I p_I-\mH^{p.g.f.}=
\nonumber \\
&=&\frac{1}{4\pi\alpha'}\left(
\frac{1}{ g_{t\xi}}
g_{IJ} \partt \bx^I
\partt \bx^J-g_{t\xi}
g_{IJ}\parts \bx^I
\parts \bx^J\right)
+ \frac{1}{2\pi\alpha'}
b_{tI}\parts \bx^I
+\frac{1}{2}
\frac{g_{tt}}{g_{t\xi}} \ .
\nonumber \\
\end{eqnarray}
It is instructive to study the
 the dynamics of the radial mode.
Using the
form of the metric components
\begin{equation}
g_{tt}=-\frac{R^2}{r^4} \ ,
\quad g_{t\xi}=\frac{R^2}{r^2} \ ,
\quad
g_{rr}=\frac{R^2}{r^2}
\end{equation}
we find that the truncanted Lagrangian
for $r$  is equal to
\begin{equation}
\mL_{r}=\frac{1}{4\pi\alpha'}(
\partt r)^2-\frac{1}{2}\frac{1}{r^2} \
\end{equation}
so that the equation of motion for $r$
takes the form
 \begin{equation}
 -\frac{1}{2\pi\alpha'}
 \partt \partt r+\frac{1}{r^3}=0 \ .
 \end{equation}
 If we multiply this equation with $\partt t$
we can rewrite it as
 \begin{equation}
 \partt
 \left[\frac{1}{4\pi\alpha'}(\partt t)^2+
 \frac{1}{2r^2}\right]=0
  \end{equation}
 and consequently
 \begin{eqnarray}\label{parttrE}
 (\partt r)^2=4\pi\alpha'\left(E-\frac{1}{2r^2}
 \right) \ , \nonumber \\
 \end{eqnarray}
 where $E>0$ is a conserved energy of the string.
The expression on the right side is positive
for
\begin{equation}
 r>\frac{1}{\sqrt{2E}} \ .
\end{equation}
It is easy to determine the solution
of the equation (\ref{parttrE})
\begin{equation}\label{rtau}
r=\frac{1}{\sqrt{2E}}
\sqrt{1+8\pi\alpha'^2 E^2 t^2} \ ,
\end{equation}
where the initial condition was
choosen such that for
 $\tau=0$ the string reaches its
turning point at $r=\frac{1}{\sqrt{2E}}$.
 In fact,
according to the equation (\ref{rtau})
the string with energy $E$ is localized
at $r=\infty$ in infinity past $\tau=-\infty$,
then moves towards to the boundary of
the space, reaches its turning point
at $r=\frac{1}{\sqrt{2E}}$ and then
moves to $r=\infty$ for $\tau=\infty$.
In other words the
 string with positive energy cannot
reach the boundary of the space at $r=0$.
Clearly this  result could have
an impact on the formulation of
the holographic relations
between non-relativistic
 quantum field
theories on the boundary and the string
modes in the bulk of the
non-relativistic background.

\section{Strings in Non-Extremal
Non-Relativistic D3-brane Background}
\label{fifth} In this section we
briefly discuss the Hamiltonian
dynamics of the string in non-extremal
non-relativistic D3-brane background
\cite{Adams:2008wt,Mazzucato:2008tr}
\begin{eqnarray}\label{nonexmet}
ds^2&=&\left(\frac{R}{r}\right)
^{2}\frac{1}{K}
\left[-\left(\frac{g}{2}+\frac{2f\triangle^2}{r^2}
\right)dt^2-\frac{g}{2}d\xi^2+(1+f)dt
d\xi+K(dx^i)^2
\right]+\nonumber \\
&+&\left(\frac{R}{r}\right)^{2}
\left[f^{-1}dr^2+r^2
\left(\frac{1}{K}(d\chi+\mathcal{A})^2+ds^2_P\right)
\right] \ , \nonumber \\
B&=&\frac{\triangle R^2}{\sqrt{2}r^2 K}
(d\chi+\mathcal{A})\wedge
((1+f)dt+(1-f)d\xi) \ ,
\nonumber \\
& & e^{\Phi}=\frac{1}{\sqrt{K}} \  ,
\nonumber \\
\end{eqnarray}
where
\begin{equation}
f=1+g \ , \quad
K=1-\beta^2 r^2 g \ , \quad
g=\frac{\rho_H^4}{r^4} \
\ ,
\end{equation}
where $\beta$ is a free parameter.
It is expected that this
background is dual to the
non-relativistic quantum field theory
at finite temperature and  with
non-zero chemical potential
\cite{Maldacena:2008wh,Adams:2008wt}.
 An important property of
this  metric (\ref{nonexmet}) is that
there is a non-zero component of
 metric $g_{\xi\xi}$ and also
non-zero component of the NS two form
field $b_{\chi \xi} \ , b_{a\xi}$.
As a consequence of this fact we
find that the components of metric
$g^{tt},g^{t\xi},g^{\xi\xi}$ are
equal to
\begin{equation}
g^{tt}=\frac{g_{\xi\xi}}{g_{tt}g_{\xi\xi}-
g_{t\xi}g_{t\xi}} \ , \quad
g^{t\xi}=g^{\xi
t}=-\frac{g_{t\xi}}{ g_{tt}g_{\xi\xi}-
g_{t\xi}g_{t\xi}} \ , \quad  g^{\xi\xi}=
\frac{g_{\xi\xi}}{g_{tt}g_{\xi\xi}-
g_{t\xi}g_{t\xi}}  \ .
\end{equation}
Then the
 Hamiltonian density is
equal to
\begin{equation}
\mH=\left(
\frac{N_T}{\sqrt{\omega}}+f_T\right)\mH_T+
(N_S+f_S)\mH_S+u_N \pi_N+u_S\pi_S+
u_\omega \pi_\omega \ ,
\end{equation}
where
\begin{eqnarray}
 \mH_S&=&p_t\parts
t+p_\xi\parts \xi+ \parts \bx^I p_I \ ,
\nonumber \\
\mH_T &=&
 \frac{1}{4\pi\alpha'} ((2\pi\alpha'
p_t-b_{tI}\parts \bx^I)g^{tt}
(2\pi\alpha' p_t-b_{tJ}\parts \bx^J)+
\nonumber \\
&+&
2(2\pi\alpha' p_\xi-b_{\xi I}\parts
\bx^I)g^{\xi t} (2\pi\alpha'p_t -b_{tJ}
\parts \bx^J)+
\nonumber \\
&+& (2\pi\alpha' p_\xi-b_{\xi I}\parts
\bx^I) g^{\xi\xi} (2\pi\alpha' p_\xi-
b_{\xi J}\parts \bx^J)+ \nonumber
\\
&+&(2\pi\alpha' p_I-b_{I\xi}\parts
\xi-b_{It}\parts t) g^{IJ}(2\pi\alpha'
p_J- b_{J\xi}\parts
\xi-b_{Jt}\parts t)+ \nonumber \\
&+& g_{tt}\parts t\parts t+2
g_{t\xi}\parts t\parts \xi+
g_{\xi\xi}\parts \xi\parts \xi+
g_{IJ}\parts \bx^I\parts \bx^J
\ . \nonumber \\
\end{eqnarray}
 Since the static gauge
can be easily formulate in the
Lagrangian formalism of given theory we
restrict ourselves to the case of the
uniform gauge defined as
\begin{equation}
G_1: t-\tau=0 \ , \quad G_2:  p_\xi-\bp=0 \ .
\end{equation}
Then we can proceed exactly in the
same way as in previous section.
Solving the constraint $\mH_S=0$
we find
\begin{equation}\label{partxi}
\parts \xi=-\frac{1}{\bp}\parts \bx^I
p_I \
\end{equation}
and solving the constraint $\mH_T=0$ for
$p_t$  we
 find the gauge fixed Hamiltonian
density $\mH^{n.b.f.}=-p_t$ in the form

\begin{eqnarray}\label{nbfH}
\mH^{n.b.f.}=
-\frac{1}{2\pi\alpha'}b_{tI}\parts
\bx^I-\frac{1}{2\pi\alpha'}\frac{g_{t\xi}}{g_{\xi\xi}}
(1-b_{\xi I}\parts \bx^I)-
\frac{1}{2\pi\alpha'
g^{tt}}\sqrt{\mathbf{K}} \ ,
\end{eqnarray}
where
\begin{eqnarray}
\mathbf{K}
&=&-\frac{1}{g_{tt}g_{\xi\xi}-g_{t\xi}^2}
\left[(1-b_{\xi I}\parts \bx^I)^2
+4\pi^2\alpha'^2g_{\xi\xi}(p_I+(\parts
\bx^Kp_K) b_{I\xi})g^{IJ}(p_J+(\parts
\bx^Lp_L) b_{J\xi})\right.
 +\nonumber \\
&+&\left.g_{\xi\xi}(\parts
\bx^Ip_I) (\parts \bx^J
p_J)+g_{\xi\xi}g_{IJ}\parts \bx^I
\parts
\bx^J\right] \ ,  \nonumber \\
\end{eqnarray}
where we set
$\bp=\frac{1}{2\pi\alpha'}$ and used
the equation (\ref{partxi}). Note  that
now the gauge fixed Hamiltonian
 takes the square-root
form as in the relativistic case which
is in contrast with the gauge fixed
Hamiltonian for string in non-relativistic
extremal background.

Clearly the same gauge fixed
Hamiltonian (\ref{nbfH})
can be used for the description
of the string in uniform gauge
that moves in
the background that is
dual to the non-relativistic quantum
field theory at zero temperature but
with non-zero particle density. This
background was derived in
\cite{Adams:2008wt} and it takes
remarkably simple form
\begin{eqnarray}
ds^2&=&\frac{1}{r^2\kappa}
(-\frac{dt^2}{r^2}+2dt d\xi+ \Omega^2
r^4 d\xi^2)+\frac{(dx^i)^2+dr^2}{r^2} \ ,
\nonumber \\
B&=&\frac{1}{r^2\kappa}(dt+2\Omega^2
r^4 d\xi)\wedge (d\chi+\mathcal{A}) \ ,
\nonumber \\
\phi&=&-\frac{1}{2}\ln \kappa \ ,
\nonumber \\
\end{eqnarray}
where for simplicity  we ignored the
sphere directions and where
$\kappa=1+\Omega^2 r^2$ and where
$\Omega$ is parameter that is related
to the particle number density in dual
field theory. This background possesses
many interesting properties and certainly
deserves to be studied further. We hope
to return to  the study of the dynamics of string
in given background in future.
\vskip 5mm
{\bf Acknowledgement:} This work was
 supported by the Czech
Ministry of Education under Contract
No. MSM 0021622409.

\end{document}